\documentclass[12pt]{article} 

\pdfoutput=1

\usepackage{amsmath, amsfonts, amssymb}
\usepackage[margin=0.75in]{geometry}
\usepackage{graphicx}
\usepackage{hyperref}
\usepackage{setspace}
\usepackage{algorithm}
\usepackage{algpseudocode}
\usepackage{tabularx}
\usepackage{lscape} 
\usepackage{rotating}
\usepackage{hyphenat} 
\usepackage{enumitem}
\usepackage{bm}
\usepackage{placeins} 
\usepackage{setspace}
\usepackage[capitalise]{cleveref}
\usepackage{fancyhdr}
\usepackage{color}
\usepackage{adjustbox} 
\usepackage{tikz} 

\usepackage{longtable} 
\usepackage{caption}

\usepackage{cite}

\usepackage{array, booktabs, makecell} 

\def \bv{\mathbf{u}}
\def \bx{\mathbf{x}}

\newcommand{\edit}[1]{\color{black}#1 \color{black}}

\title{Uncovering near-wall blood flow from sparse data with physics-informed neural networks }
\author{Amirhossein Arzani$^1$ \and Jian-Xun Wang$^2$ \and Roshan M. D'Souza$^3$ }

\date{}

\begin{document}

\maketitle


\begin{center}
$^1$Department of Mechanical Engineering, Northern Arizona University, Flagstaff, AZ, United States \\
$^2$Department of Aerospace and Mechanical Engineering, University of Notre Dame, Notre Dame, IN,  United States \\
$^3$Department of Mechanical Engineering, University of Wisconsin--Milwaukee, Milwaukee, WI, United States.\\
\end{center}

\bigskip

\noindent Correspondence:\\
Amirhossein Arzani,\\
Northern Arizona University,\\
Flagstaff, AZ, 86011\\
Email: amir.arzani@nau.edu

\thispagestyle{empty}



\begin{abstract}

Near-wall blood flow and wall shear stress (WSS) regulate major forms of cardiovascular disease, yet they are challenging to quantify with high fidelity. Patient-specific computational and experimental measurement of WSS suffers from uncertainty, low resolution, and noise issues.  Physics-informed neural networks (PINN) provide a flexible deep learning framework to integrate mathematical equations governing blood flow with measurement data. By leveraging knowledge about the governing equations (herein, Navier-Stokes), PINN overcomes the large data requirement in deep learning. In this study, it was shown how PINN could be used to improve WSS quantification in diseased arterial flows. Specifically, blood flow problems where the inlet and outlet boundary conditions were not known were solved by assimilating very few measurement points. Uncertainty in boundary conditions is a common feature in patient-specific computational fluid dynamics models. It was shown that PINN could use sparse velocity measurements away from the wall to quantify WSS with very high accuracy even without full knowledge of the boundary conditions. Examples in idealized stenosis and aneurysm models were considered demonstrating how partial knowledge about the flow physics could be combined with partial measurements to obtain accurate near-wall blood flow data. The proposed hybrid data-driven and physics-based deep learning framework has high potential in transforming high-fidelity near-wall hemodynamics modeling in cardiovascular disease.




\end{abstract}

\newpage

\section{Introduction} \label{sec:intro}

There have been tremendous advances in physics-based modeling of patient-specific cardiovascular flows~\cite{VardhanRandles21}. These models improve our understanding of blood flow dynamics (hemodynamics) in diseased arteries; however, translating these models into clinical practice is a challenging task. Major forms of cardiovascular disease such as atherosclerosis and thrombosis localize at the vessel wall. We have learned that fundamental fluid dynamics processes regulate cardiovascular disease near the wall~\cite{Arzanietal16,Mahmoudietal20}, and clinical studies demonstrate the correlation between near-wall blood flow parameters such as wall shear stress (WSS) and cardiovascular disease~\cite{Samadyetal11,Detmeretal20}. In the past three decades, near-wall hemodynamics have received the utmost attention in cardiovascular fluid mechanics research. 

Unfortunately, characterization and quantification of near-wall flow and WSS is challenging. The complex and disturbed (often chaotic~\cite{ArzaniShadden12} and turbulent~\cite{AntigaSteinma09})  blood flow dynamics in diseased arteries translate to complex WSS patterns on the vessel wall~\cite{ArzaniShadden18}. WSS vectors represent rich features with spatial and temporal variations in their magnitude and direction~\cite{ArzaniShadden16}. Additionally, the rich topological structures in the WSS vector field~\cite{ArzaniShadden18,Mazzietal21} and their strong role in controlling near-wall mass transport~\cite{Arzanietal16,Mahmoudietal20} further complicate its characterization. All of these complex patterns in WSS increase its sensitivity to modeling assumptions and limitations. Perhaps the simplest way to see this is from Poiseuille's law, where WSS has a strong dependence on variations in diameter ($\tau \sim 1/ D^3$). Computational fluid dynamics (CFD) quantification of WSS is sensitive to uncertainty in image segmentation (3D geometry) as well as uncertainty in boundary conditions and blood rheology. In-vitro experimental measurement of WSS, e.g.,\ particle image velocimetry (PIV), has the same uncertainty issues as CFD modeling. Additionally, PIV measurements near the wall are challenging due to inherent limitations in the PIV technique~\cite{Raffeletal18,Wangetal20}.  Collectively, these challenges have prevented the widespread clinical use of WSS. 

Recent advances in data-driven modeling  have shown promise for transforming cardiovascular flow research and addressing challenges in hemodynamics modeling~\cite{ArzaniDawson21}. Specifically, different types of data assimilation algorithms have been proposed to merge experimental and computational hemodynamics data to improve data fidelity~\cite{Bakhshinejadetal17,Gaidziketal19,Funkeetal19,ArzaniDawson21,Habibietal21,GaoWang21}. Discovering a reduced-order basis with unsupervised machine learning (e.g., principal component analysis) is typically a key step in these methods to either reduce the computational cost~\cite{Habibietal21} or enable a compressed sensing optimization problem~\cite{ArzaniDawson21}. These studies have shown promise in improving near-wall hemodynamics prediction~\cite{Habibietal21}. A typical limitation of some of these methods is the a priori estimation of modeling error/uncertainty, which are usually not known in experimental measurements and are computationally expensive to estimate for computational models. Traditional data assimilation methods usually require a large number of repeated forward evaluations of expensive CFD models or intrusive implementation of an adjoint solver, which is often computationally expensive and less scalable. 


Machine learning and deep learning are other recent paradigms that could improve the fidelity in hemodynamics models. Specifically, scientific machine learning deals with problems where data are sparse, incomplete, and sometimes have low fidelity. These are pertinent features in patient-specific hemodynamics modeling. For example, in 4-dimensional (4D) flow magnetic resonance imaging (MRI), we have low resolution and noisy data, or when modeling blood flow with PIV the signal-to-noise ratio near the wall could be lower than the core flow region. In general, a machine learning framework that can handle the scarcity of data is highly desirable in modeling hemodynamics. Physics-informed neural networks (PINN) have been recently proposed, where the governing physical equations (e.g., incompressible Navier-Stokes that govern continuum scale blood flow) could be leveraged in the neural network training~\cite{Raissietal19}. Namely, we could design a neural network that is trained to satisfy the given training data as well as the imposed governing equations. This is a powerful framework once we recall that neural networks are universal function approximators~\cite{Horniketal89}, and therefore could approximate spatiotemporal hemodynamic data. \edit{PINN has been applied to a wide range of complex fluid flow problems~\cite{LeeYou19,Wesselsetal20,Kashinathetal21,Lucoretal21,GasmiTchelepi21}.}

In general, deep neural networks could approximate any high-dimensional function once sufficient training data are supplied. However, an interesting feature of PINN is that we do not necessarily need any training data in solving a well-defined problem~\cite{Jinetal21}. That is, PINN could find velocity and pressure fields that satisfy Navier-Stokes over arbitrarily specified points in the domain and we may think of this as a meshfree alternative to traditional CFD approaches~\cite{Luetal21}. This paradigm has been applied to different fluid flow problems~\cite{Hennighetal20}. However, replacing CFD with a neural network that at best can solve a given set of equations  does not address the issues raised above in high-fidelity modeling of hemodynamics. The flexibility in defining any given problem is what makes PINNs powerful. For instance, we may incorporate the governing equations without knowing the boundary conditions, which is typically the case in patient-specific modeling. Additionally, we may guide the neural network with training data that do not necessarily need to be large and also learn the solution of differential equations in a parametric setting (surrogate modeling). With some knowledge about the physics of our problem (even with uncertainty) and some form of training data (even sparse and incomplete), we may have hope to use PINN for finding an optimal solution with high fidelity.  

The flexible, hybrid, data-driven, and physics-based framework that PINN offers has recently enabled solutions to complicated cardiovascular biomechanics problems. For example, PINN has been used for fast surrogate modeling of idealized blood flow problems in a forward parametric setting without labeled training data~\cite{Sunetal20}. Moreover, PINN has also been formulated in an inverse modeling setting to obtain blood flow velocity data from time-resolved concentration data in aneurysms~\cite{Raissietal20,Caietal21}. Pressure estimation from low-fidelity 4D flow MRI velocity data has been performed using PINN and 1D blood flow models~\cite{Kissasetal20}.  PINN could also be used to learn constitutive laws for non-Newtonian fluids~\cite{Reyesetal20} or tissue material property~\cite{LiuLiangSun20,Buosoetal21}, which are useful in modeling multiphase and multiphysics cardiovascular flows. 

An interesting application of neural networks is superresolution of low-resolution fluid flow data. Direct numerical simulation data have been used to train neural networks for superresolution of low-resolution turbulence data~\cite{Liuetal20,Guemesetal21,Fukamietal21}. PINN has been used to eliminate or reduce the need for training data during superresolution~\cite{GaoSunWang20} and has been applied to low resolution and noisy blood flow data derived from 4D flow MRI~\cite{Fathietal20}. In most of these studies, the input low-resolution data is distributed throughout the entire region of interest. In this study, we would like to obtain high-resolution blood flow data from very few sensors not distributed throughout the entire region. Specifically, we are interested in quantifying WSS from sparse sensor measurements primarily away from the wall where the sensors do not cover the entire region of interest. Previous work has used neural networks to quantify velocity from WSS measurements~\cite{Guastonietal20} or obtain WSS from PIV measurements where the near-wall data is not accurate~\cite{Wangetal20}. However, these studies used CFD training data in building their model, which could be computationally expensive. 

The goal of the present work is to obtain accurate WSS data from sparse velocity measurements away from the wall without any training data. In our problem, the inlet and outlet boundary conditions are treated as unknowns, which is a common scenario in patient-specific hemodynamics modeling. Our problem is therefore not mathematically well-posed and could not be solved using traditional CFD solvers. Dealing with sparse, incomplete data and uncertain mathematical models is a key feature of multiscale and predictive modeling of cardiovascular systems~\cite{Pengetal20}, where WSS plays an important role by linking blood flow to the biological events at the cell-scale~\cite{Mahmoudietal20}.  We will demonstrate different examples of how PINN could be used to obtain WSS from sparse measurements and uncertain mathematical models (i.e.,\ given some data and some physics).

\section{Methods} \label{sec:meth}
\subsection{Problem statement}
Consider the steady, incompressible, Newtonian, Navier-Stokes equations 

\begin{subequations} \label{eq:NS}
\begin{equation}
 \rho \bv  \cdot \nabla \bv = -\nabla p + \mu   \nabla^2 \bv  
\end{equation}
\begin{equation}
\nabla \cdot \bv = 0 \;,
\end{equation}
\end{subequations}
where $\bv(\bx)$ is the velocity vector, $p(\bx)$ is the pressure field, $\bx \in \Omega$ is the spatial coordinate, $\mu$ is the dynamic viscosity, and $\rho$ is the density. Boundary conditions (BC) could be provided as
\begin{equation}
\mathcal{B}(\bx,p,\bv) = 0 \; \; \; \bx \in \partial\Omega \;,
\end{equation}
where $\mathcal{B}$ is an appropriately defined differential operator that determines admissible boundary conditions on the boundary $ \partial\Omega$. If we precisely know the parameters in Navier-Stokes ($\mu$ and $\rho$) and the boundary condition $\mathcal{B}$, then the above equation could be solved using traditional CFD methods. However, in patient-specific cardiovascular fluid mechanics applications, the parameters and BCs are either partially known or they are associated with uncertainty.

Suppose that we are not given the parameters (e.g.,\ $\mu$) or we do not know the boundary conditions on the entire boundary $ \partial\Omega$. A common scenario is when $\mathcal{B}$ is only known on the vessel wall (no-slip, no-penetration) but is missing at the inlet or outlets.  Obviously, we can no longer solve this problem using traditional numerical methods. The issue is that now the problem is not well-posed and therefore we do not have a unique solution.  However, we propose that given sparse measurements in the domain $\Omega$, we may have hope to find the solution. Assume that the velocity vector field is measured in a sparse set of measurement points
\begin{equation}
\bv(\bx_i) = \mathbf{f}(\bx_i) \; \; \; \bx_i \in \Gamma \;,
\label{eqn:data}
\end{equation}
where $\Gamma =\{\bx_i\} $, i=1, $\cdots$, N, is a set of N discrete measurement locations where the velocity is given by the function $\mathbf{f}$. Our goal is to find the best solution $\bv(\bx)$ that satisfies these measurement data and also satisfies the Navier-Stokes equations with partially defined boundary conditions or unknown parameters. Specifically, in this study, we are interested in leveraging sparse measurement points away from the wall to obtain near-wall flow and WSS.

\subsection{Physics-informed neural networks (PINN)}
Neural networks (NN) are a powerful class of machine learning algorithms that are capable of approximating complex, high-dimensional, and nonlinear functions. In this work, we are interested in using NNs to approximate the velocity vector as a function of space, i.e.,\ $\bv(\bx)$. Specifically, fully-connected NNs are used in this study, where a composite function is formed by stacking together layers of artificial neurons where each layer $\mathbf{z}_k$ performs an affine transformation on the previous layer with nonlinearity imposed via an element-wise activation function $\mathbf{\sigma}$
\begin{equation}
\mathbf{z}_k = \mathbf{\sigma}(\mathbf{W}^T_{k-1} \mathbf{z}_{k-1} + \mathbf{b}_{k-1})  \;,
\end{equation}
where $\mathbf{W}_k$ and  $\mathbf{b}_k$ are the weight matrices and bias vectors for each layer k. The weights and biases are obtained by a variant of stochastic gradient descent optimization (herein, ADAM). Swish activation function~\cite{Ramachandranetal17} \edit{($ x \times Sigmoid(\beta x)$ with $\beta$=1)}  is used for all layers \edit{and all cases.} Velocity vector components and the pressure are each approximated with a separate NN. In PINN, we leverage our prior knowledge of the physics of the problem to obtain a loss function for the optimization problem. Specifically, the Navier-Stokes equations are used as the physics loss function $\mathcal{L}_{phys} $
\begin{equation}
\mathcal{L}_{phys} = \left\lVert  \rho \bv  \cdot \nabla \bv +\nabla p - \mu  \nabla^2 \bv  \right\rVert_\Omega +  \left\lVert \nabla \cdot \bv    \right\rVert_\Omega   \;.
\end{equation}
The boundary conditions are enforced using a boundary condition loss  $\mathcal{L}_{BC} $
\begin{equation}
\mathcal{L}_{BC} =  \left\lVert  \mathcal{B}(\bx,p,\bv)  \right\rVert_{\partial\Omega}  \;.
\end{equation}

\begin{figure}[h!]
\centering
\includegraphics[width=.8\textwidth]{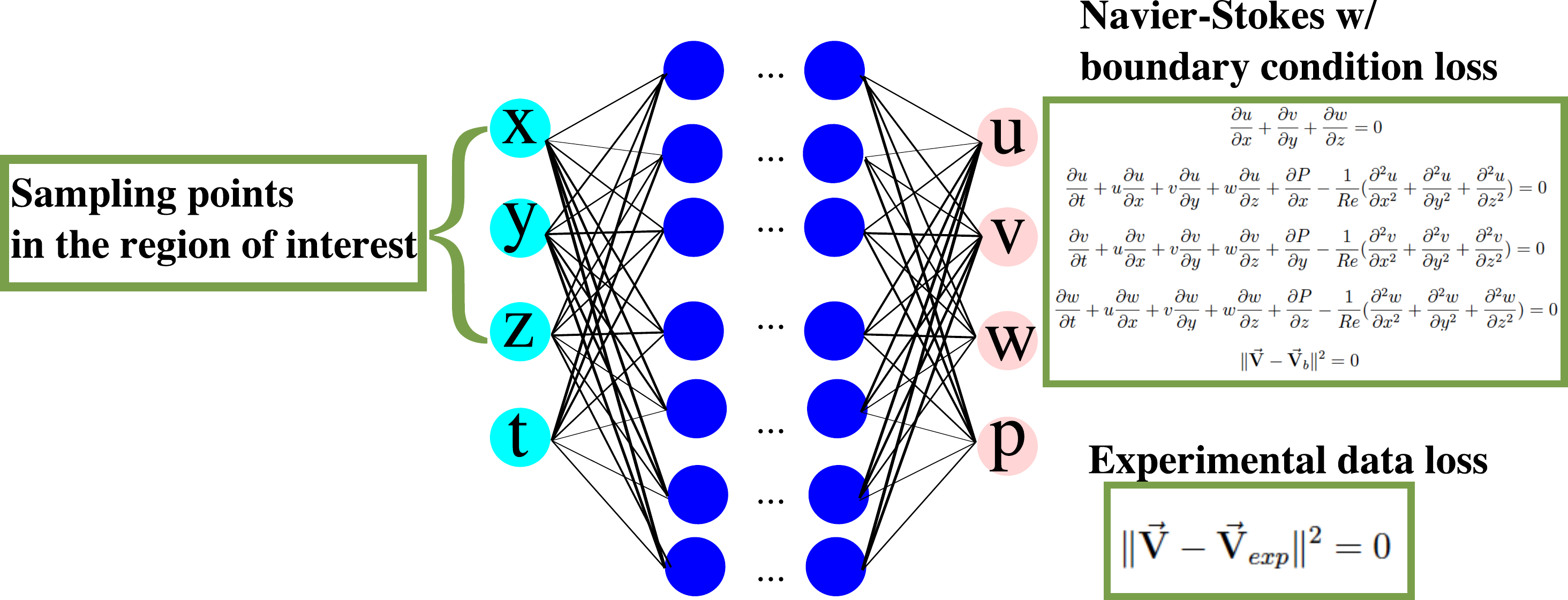}
\caption{Physics-informed neural networks (PINN) could be used for solving Navier-Stokes equations. PINN could be augmented with experimental measurement data to improve the predicted solution and/or eliminate the necessity of fully defining boundary conditions or parameters. The spatial input points in PINN represent a flexible and meshfree representation of the domain. } 
\label{fig:pinn}
\end{figure}

With an appropriate definition of the boundary conditions $ \mathcal{B}$, the above two objectives (Navier-Stokes and BC) provide a mathematically well-posed problem, which could be solved with PINN even without training data~\cite{Jinetal21}. However, in patient-specific cardiovascular modeling, the BCs are often not fully known. Therefore, in this study, $\mathcal{L}_{BC} $ is only enforced on parts of the boundary. In traditional numerical methods, solving partial differential equations (PDEs) without well-posed BCs is not always feasible; however, PINN is able to converge to a solution that minimizes the governing equation's residual and the partially known (or fully unknown) BC. Nonetheless, this will likely not lead to the true solution as there are infinite possibilities for solving a PDE without fully imposed BCs. To mitigate this issue, we leverage sparse data measurements to define a new data loss $\mathcal{L}_{data} $
\begin{equation}
\mathcal{L}_{data} =  \left\lVert \mathbf{z} - \mathbf{f}    \right\rVert_{ \Gamma}  \;,
\end{equation}
where $ \mathbf{f} $ is given by Eq.~\ref{eqn:data} and $\mathbf{z}$ represents the velocity measurements at sparse locations $ \Gamma$. The total loss function is defined by a linear combination of the above loss functions
\begin{subequations} \label{eq:ml}
\begin{equation}
\mathcal{L}_{tot}(\mathbf{W}_i, \mathbf{b}_i ) = \mathcal{L}_{phys} + \lambda_{b} \mathcal{L}_{BC} +  \lambda_{d} \mathcal{L}_{data} \;,
\end{equation}
\begin{equation}
\mathbf{W}_i^{*}, \mathbf{b}_i ^{*} = \operatorname*{arg\,min}_{\mathbf{W}_i, \mathbf{b}_i}  \mathcal{L}_{tot}(\mathbf{W}_i, \mathbf{b}_i )  \;,
\end{equation}
\end{subequations}
where the optimal weights and biases for each velocity component and pressure network $\mathbf{W}_i^{*}$ and $\mathbf{b}_i ^{*} $ are obtained such that the total loss function $\mathcal{L}_{tot}$ is minimized. The hyperparameters  $ \lambda_{b}$ and $ \lambda_{d}$ are used to weight the contributions of BCs and measurement data, respectively. The mean squared error (MSE), equivalently the squared L2 norm, is used in computing the losses. An overview of PINN is shown in Fig.~\ref{fig:pinn}.

\subsection{Blood flow problems}
In this Section, we describe different test cases and the computational fluid dynamics (CFD) simulations that were performed and treated as ground truth data. Subsequently, we define the problem where given sparse measurement sensors and incomplete boundary conditions we would like to quantify near-wall blood flow and WSS using PINN. All CFD simulations were performed using FEniCS~\cite{LoggMardalWells12}, which is an open-source finite element solver. PINN models were implemented in Pytorch. All geometric dimensions and variables were normalized, which facilitates convergence in NNs. In all numerical problems, the same nodes in the CFD mesh were used as sampling points in PINN. The inlet/outlet BC was assumed unknown in the PINN problem and the best solution satisfying the Navier-Stokes equations, the no-slip BC, and the sparse measurements (sampled from the CFD data) was found.  

\subsubsection{Test case 1: 1D advection-diffusion transport}
First, we consider a simple 1D advection-diffusion equation with an available analytical solution to demonstrate the feasibility of our method. Convective mass transport equations are commonly solved in cardiovascular mass transport models of atherosclerosis and thrombosis~\cite{Mahmoudietal20,LeidermanFogelson11} and represent numerical challenges due to large Peclet and Schmidt numbers, which result in thin concentration boundary layers~\cite{Arzanietal16,HansenArzaniShadden19}. We would like to find the wall concentration on a boundary with a thin boundary layer without any given boundary conditions.

Consider the steady advection-diffusion equation
\begin{equation}
a \frac{\partial c}{\partial x} = \nu \frac{\partial^2 c}{\partial x^2} \;,
\end{equation}
where c(x) is concentration,  $a >$ 0 is a given velocity, and $\nu$ is the diffusion coefficient. We would like to solve this problem on  $x \in$ [0,1] . Given boundary conditions on both ends, i.e.\ $c(0) = c_1 $ and $c(1) = c_2 $, we may obtain an analytical solution
\begin{equation}
c(x) = \frac{c_2 - c_1} { e^{\frac{a}{\nu}} -1 } e^{\frac{a}{\nu}x} + \frac{c_1 e^{\frac{a}{\nu}} - c_2 } {  e^{\frac{a}{\nu}} - 1} \;.
\label{eqn:analy}
\end{equation}

In this example, we set $c_1 = 1.0 $, $c_2 = 0.1$, $a=1$, and $\nu=0.01$ to generate ground truth data using the above analytical solution. Subsequently, three measurement points were collected where two points were outside of the boundary layer and one point was inside the sharp boundary layer. The values of $c_1$ and $c_2$ (BCs) are unknown and we would like to solve this problem to find the unknown boundary conditions and also c(x). 100 uniformly spaced sampling points and $\lambda_d = 10 $ were used in PINN without any BC enforcement. The network had 10 hidden layers with 100 neurons per layer to capture the thin boundary layer. A dynamically adaptive learning rate was adopted with the learning rate varying between 1e-3 and 1e-6 with a \edit{step} decay over 5000 epochs.

\subsubsection{Test case 2: 2D blood flow in a stenosis}
\begin{figure}[h!]
\centering
\includegraphics[width=\textwidth]{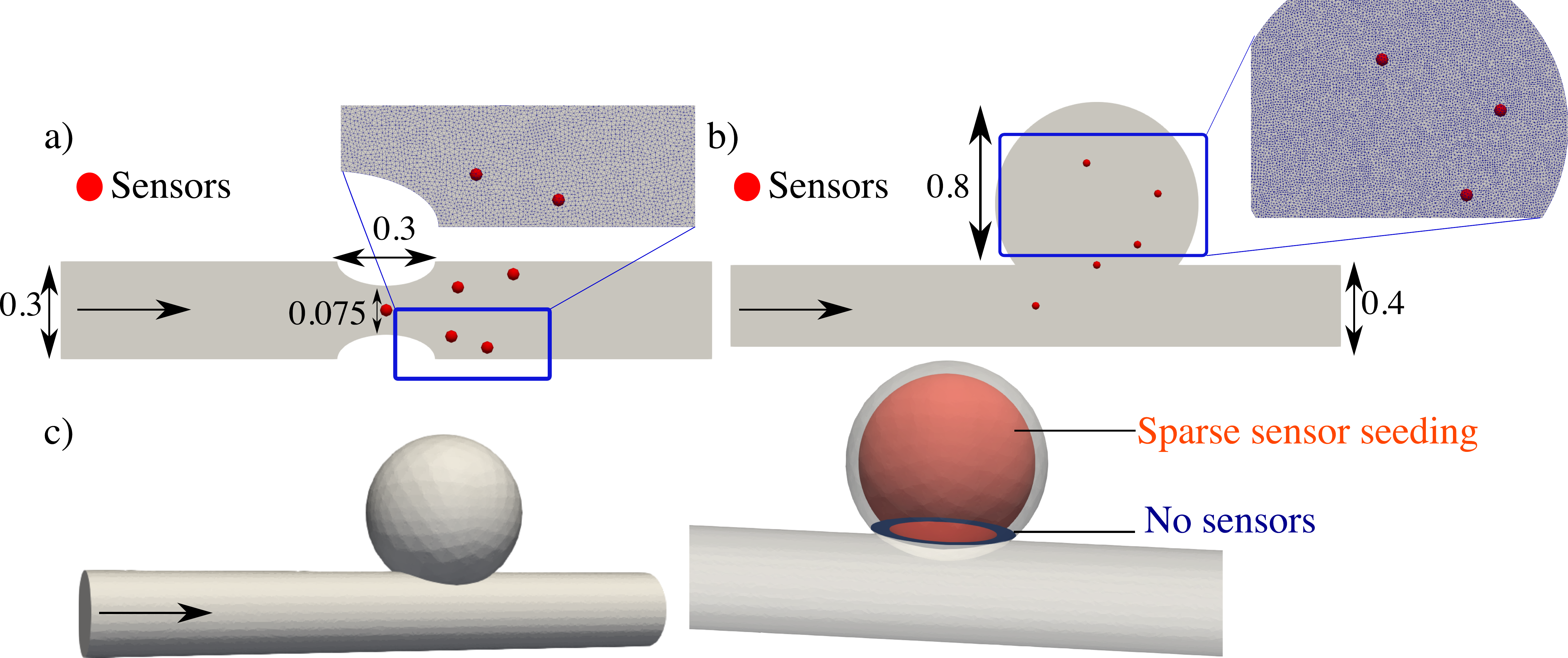}
\caption{The 2D and 3D models used in this study are shown. The red dots show the sensor locations (5 point measurements) in the 2D problems and the sensors in the 3D problem were selected on the red surface shown. Zoomed-in view of the computational mesh is used to compare the resolution in the high-resolution CFD models and the sparse measurements used by PINN. The arrows show the direction of the incoming flow.  a) The 2D stenosis, c) 2D aneurysm, and c) 3D aneurysm models are shown. \edit{In the 3D model, no boundary condition was applied at the blue plane marked as ``No sensors''.} All geometric lengths are dimensionless. The 3D aneurysm model had the same dimensions as the 2D aneurysm model.} 
\label{fig:geo}
\end{figure}

We consider an idealized 2D blood flow problem in a stenosed artery (vessel constriction). The geometric dimensions and sensor locations (5 sensors) are shown in Fig.~\ref{fig:geo}a. To generate the ground truth data, a parabolic velocity profile with a Reynolds number of Re = 150 (based on peak velocity) was prescribed at the inlet, no-slip BCs were applied at the walls, and zero traction was used at the outlet. The Re number corresponds to blood flow in coronary arteries~\cite{ShaddenHendabadi13} where near-wall hemodynamics around the plaque play an important role in regulating disease growth~\cite{Arzani20}. The mesh consisted of  41k quadratic triangular elements (corresponding to $\sim$164k linear elements). In the PINN solution, the inlet and outlet BCs were treated as unknowns and the sparse sensor measurements from ground truth data were used to guide the solution. The neural networks used in approximating pressure and velocity components had 9 hidden layers with 128 neurons per layer. The learning rate varied between 5e-4 and 5e-8 with a  \edit{step} decay over 5500 epochs. $\lambda_d = 1 $ and $\lambda_b = 20 $ were selected in weighting the loss functions and a batch size of 256 was used.

\subsubsection{Test case 3: 2D blood flow in an aneurysm}
We consider an idealized 2D blood flow problem in an idealized saccular aneurysm problem (sudden expansion). The geometric dimensions and sensor locations (5 sensors) are shown in Fig.~\ref{fig:geo}b. Similar boundary conditions were used in this problem with Re = 320. The Re number corresponds to blood flow in cerebral aneurysms formed in the internal carotid artery~\cite{Baeketal10} where hemodynamics in the aneurysm region is of clinical interest~\cite{Cebraletal11}.  The mesh consisted of  61k quadratic triangular elements (corresponding to $\sim$244k linear elements). Similar to the previous case, the inlet and outlet boundary conditions were treated as unknown and PINN was used to find velocity and subsequently WSS. The neural networks used had the same architecture as the previous example. All parameters in solving the NN were selected similar to the previous example except $\lambda_b = 1 $. 

\subsubsection{Test case 4: 3D blood flow in an aneurysm}
The previous 2D aneurysm model was extended to 3D. The geometry is shown in Fig.~\ref{fig:geo}c. Similar boundary conditions as the 2D aneurysm model were used. The mesh consisted of 1M quadratic tetrahedral elements (corresponding to $\sim$8M linear elements). In this test case, a slightly different problem was defined. Namely, a near-wall region was defined between the aneurysm wall and a surface with a normal-to-wall distance of 0.13R where R is the aneurysm radius. Measurement sensors were placed on the outer boundary of this near-wall region (red surface in Fig.~\ref{fig:geo}c). These measurement points were randomly sampled such that the total number of sensors was 200 times coarser than the number of surface nodes formed by the intersection of the red surface in Fig.~\ref{fig:geo}c and the 3D CFD mesh, \edit{which resulted in 125 sensors.} The inlet/outlet boundary condition (blue plane) in Fig.~\ref{fig:geo}c \edit{where no sensors were placed} was treated as unknown and PINN simulation was performed for the CFD nodes in the near-wall region. The neural networks used in approximating pressure and velocity had 9 hidden layers with 200 neurons per layer. The adaptive learning rate varied between 5e-4 and 5e-7 with a  \edit{step} decay over 8500 epochs. $\lambda_d = 20 $ and $\lambda_b = 20 $ were selected in weighting the loss functions and a batch size of 512 was used.

Note that imposing velocity boundary conditions on the outer wall surface (red surface) and the inlet/outlet surface (blue surface) defines a mathematically well-posed problem that could be solved with traditional methods. However, the current problem challenges traditional approaches in multiple ways. First, we do not have the boundary condition on the outer wall defined for all points. Second, the inlet and outlet boundary conditions are not known. Finally, we do not need to design a 3D mesh that is specifically created for the near-wall region. We will demonstrate how we can solve this problem with PINN to obtain WSS. Specifically, this example demonstrates the flexibility of PINNs in handling a subset of a 3D domain as the region of interest for finding the numerical solution.

\subsubsection{Test case 5: Parameter identification (viscosity)}
The previous examples focused on finding the solution when the inlet and outlet boundary conditions were unknown. Other essential parameters in solving the Navier-Stokes equations could alternatively be made unknown. In this example, the viscosity ($\mu$) was treated as unknown and the 2D stenosis problem above was repeated to identify the unknown viscosity in an inverse problem. Variations in viscosity \edit{occurring in hemorheological disorders (e.g., diabetes)~\cite{ChoMooneyCho08}} and potential non-Newtonian effects are a common problem in cardiovascular flow modeling~\cite{Arzani18}. The viscosity was assumed to be constant in space, i.e.\, we assumed Newtonian flow similar to the CFD simulations; however, spatial dependence could easily be considered in the framework. \edit{The learning rate for viscosity was ten times smaller than the velocity and pressure learning rates.}

\section{Results} \label{sec:Res}

\subsection{Test case 1: 1D advection-diffusion transport}

\begin{figure}[h!]
\centering
\includegraphics[scale =0.5]{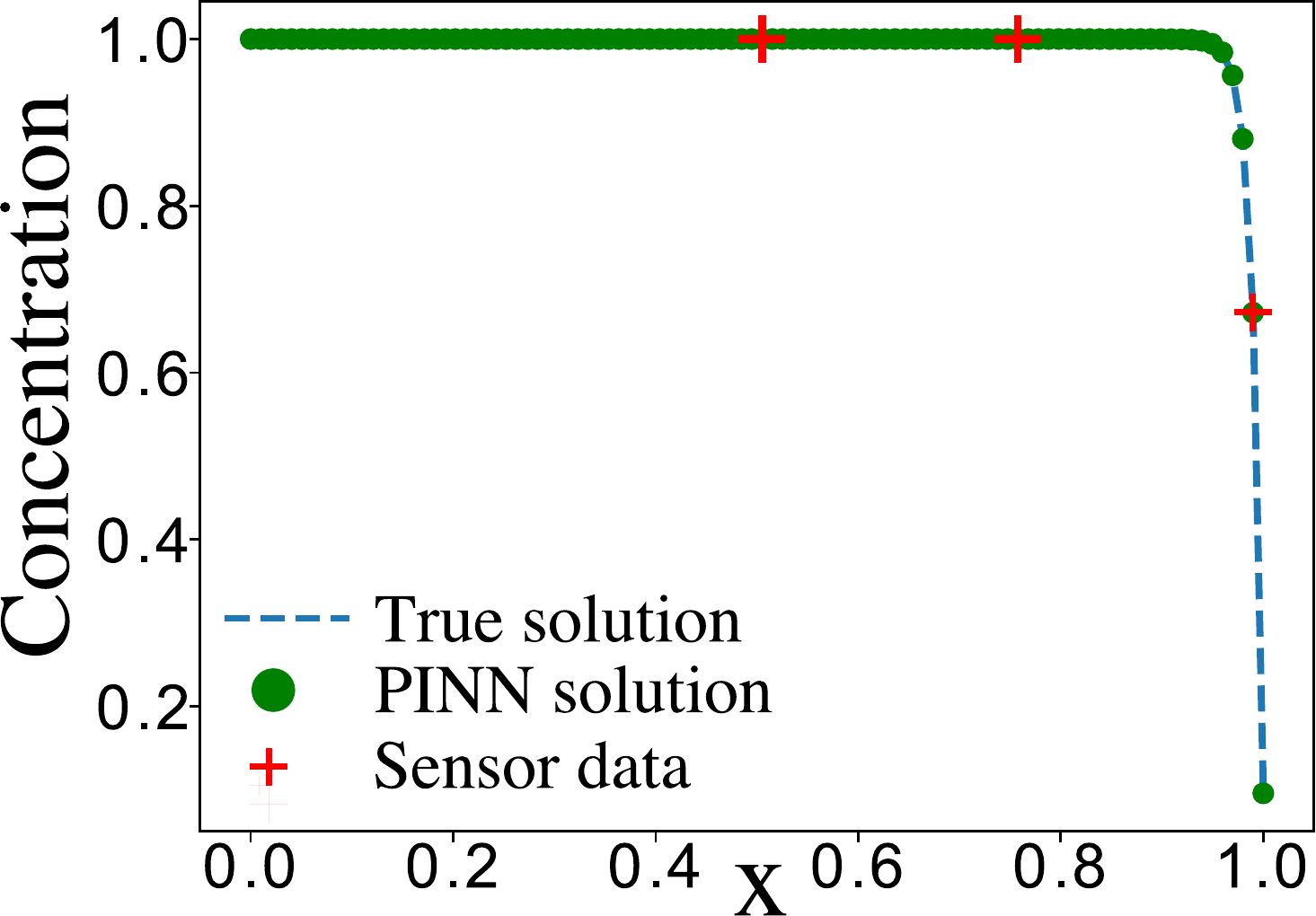}
\caption{The 1D advection-diffusion results are plotted. The three red crosses show the locations where data sampling was performed. The true solution (analytical solution) and the PINN prediction (with unknown boundary conditions) match very well.   } 
\label{fig:1d}
\end{figure}

The solution for the 1D advection-diffusion problem is plotted in Fig.~\ref{fig:1d}. The three measurement points are shown with red cross marks. The PINN solution matches the analytical solution very well and is able to identify the solution at the boundaries with unknown BCs even with a thin boundary layer.

\subsection{Test case 2: 2D blood flow in a stenosis}

\begin{figure}[h!]
\centering
\includegraphics[width=\textwidth]{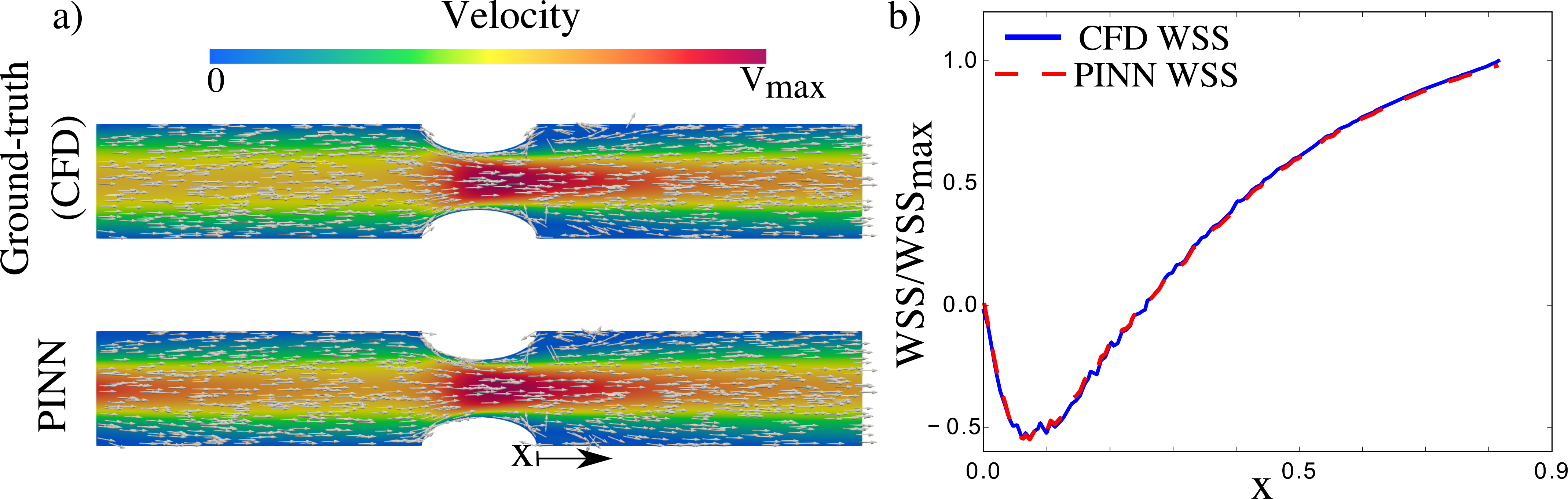}
\caption{a) The ground truth (CFD) and PINN velocity results in the 2D stenosis model are shown. Normalized velocity vectors are plotted to show the flow direction. b) The wall shear stress (WSS) results in the region distal to the stenosis are compared between the CFD and PINN models.  The inlet and outlet boundary conditions were not specified in the PINN problem. } 
\label{fig:sten}
\end{figure}

The results for the second test case (2D stenosis) are shown in Fig.~\ref{fig:sten}. The velocity contour results for the ground truth CFD simulation and the PINN prediction are shown. Normalized vectors showing the velocity direction are superimposed on the contours. An interesting observation is that PINN is unable to identify the correct inlet boundary condition; however, the velocity results in the stenosed region where sensors were placed match between the PINN and CFD simulations. The WSS results are plotted in the region distal to the stenosis \edit{where more complex blood flow patterns are observed and most sensors are placed.}  An excellent match between the CFD and PINN WSS results could be seen.

\subsection{Test case 3: 2D blood flow in an aneurysm}

\begin{figure}[h!]
\centering
\includegraphics[width=\textwidth]{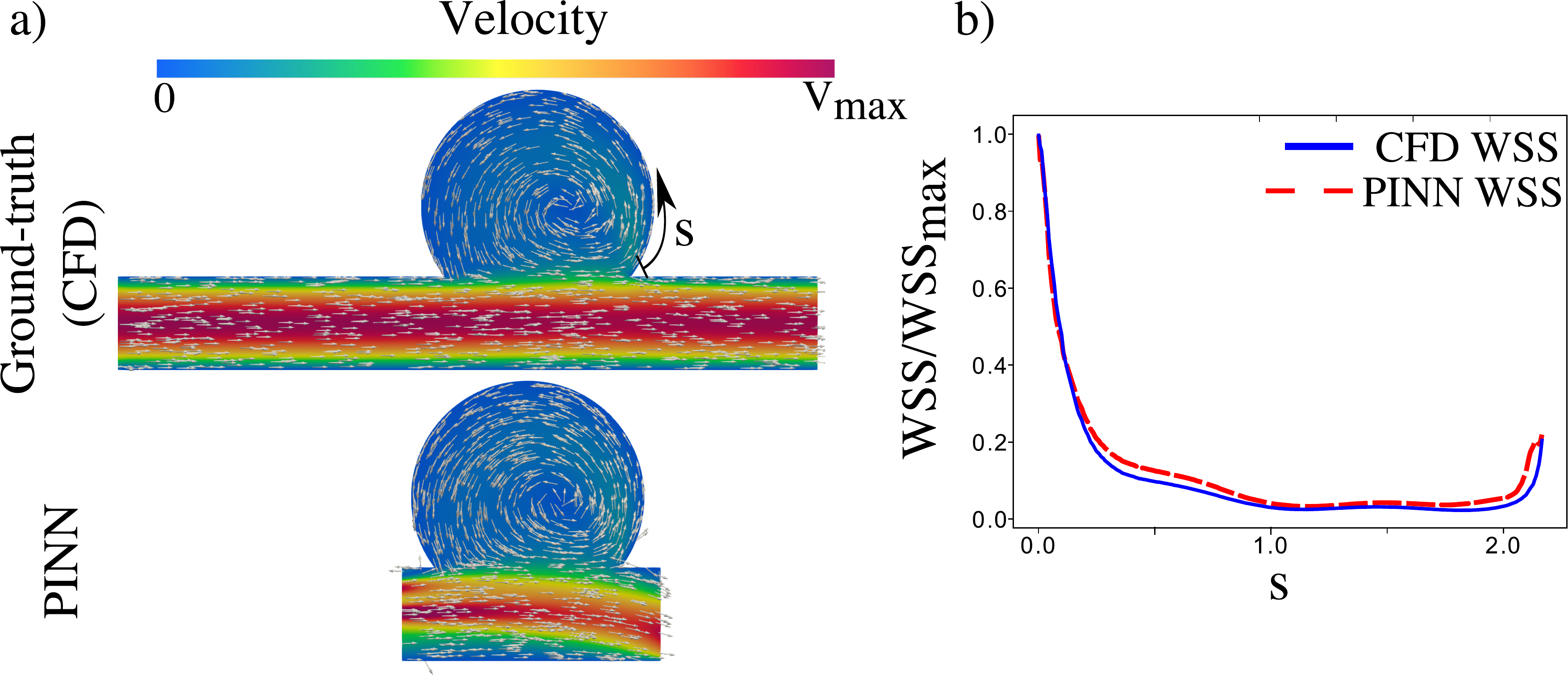}
\caption{a) The ground truth (CFD) and PINN velocity results in the 2D aneurysm model are shown. Normalized velocity vectors are plotted to show the flow direction. b) The wall shear stress (WSS) results in the aneurysmal region are compared between the CFD and PINN models.  The inlet and outlet boundary conditions were not specified in the PINN problem.  } 
\label{fig:ia2d}
\end{figure}

The 2D aneurysm model results are shown in Fig.~\ref{fig:ia2d}. In this example, PINN simulation was only performed in a cropped domain as shown in the figure, demonstrating the flexibility in PINN for selecting the region of interest. This is particularly useful here because we do not know the inlet BC. Results similar to the stenosis model could be seen where PINN does not recover the correct incoming flow pattern, however, it is capable of accurately reconstructing the flow in the aneurysm sac where sensors were placed. The WSS results on the aneurysm wall are plotted demonstrating excellent agreement with the ground truth CFD data. 

\subsection{Test case 4: 3D blood flow in an aneurysm}

\begin{figure}[h!]
\centering
\includegraphics[width=\textwidth]{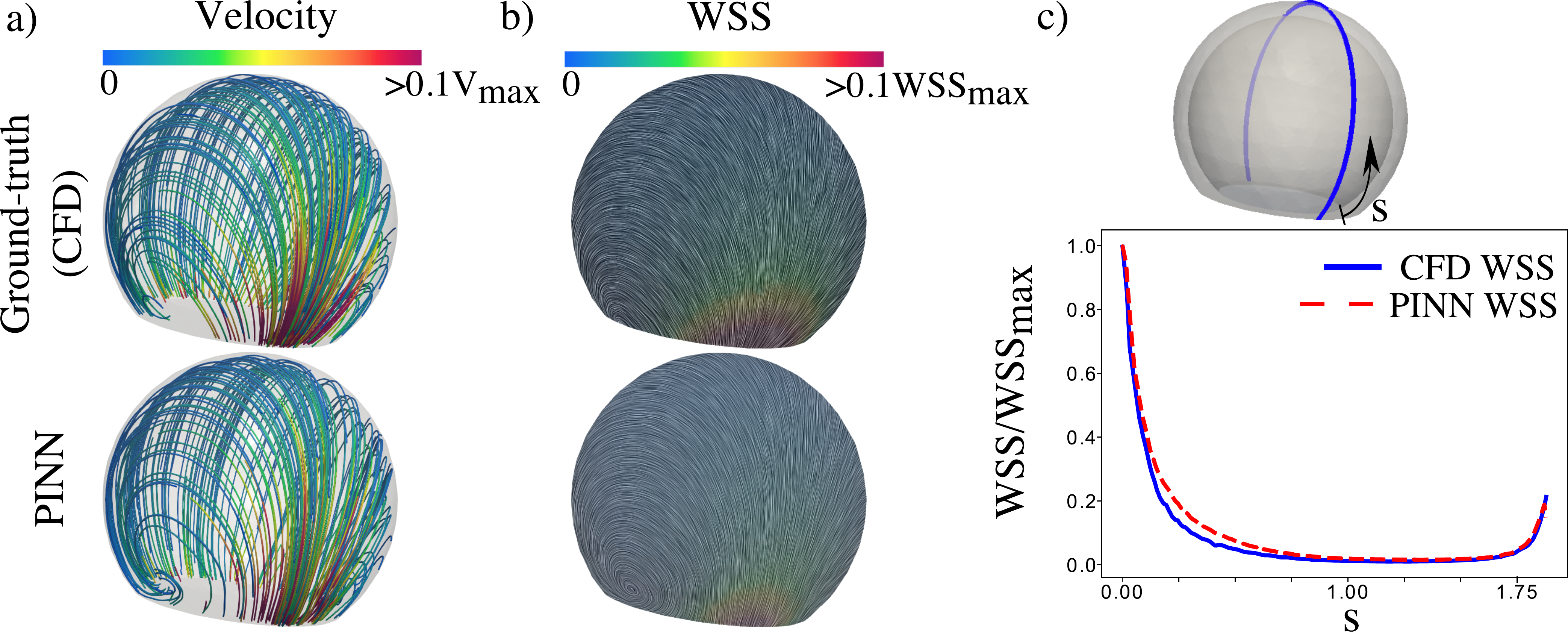}
\caption{a) The ground truth (CFD) and PINN velocity streamlines colored by their magnitude are shown for the 3D aneurysm model. b) The wall shear stress (WSS) streamlines (limiting streamlines, skin friction lines) colored by their magnitude are shown for the CFD and PINN models. c) WSS is plotted on the shown blue line and compared between the CFD and PINN models. The inlet and outlet boundary conditions were not specified in the PINN problem.   } 
\label{fig:ia3d}
\end{figure}

The 3D aneurysm velocity and WSS results are shown in Fig.~\ref{fig:ia3d}. The velocity and WSS streamlines (limiting streamlines) colored by their magnitude are compared between the CFD and PINN models. For a better quantitative comparison, WSS is plotted over the blue line in Fig.~\ref{fig:ia3d} (center of the aneurysm). The agreements between PINN and ground truth CFD results are very good, however, at some locations, a small error could be observed, which was not previously seen in the 2D results. Nevertheless, the PINN prediction has very good accuracy.
 
\subsection{Convergence of losses}

\begin{figure}[h!]
\centering
\includegraphics[width=\textwidth]{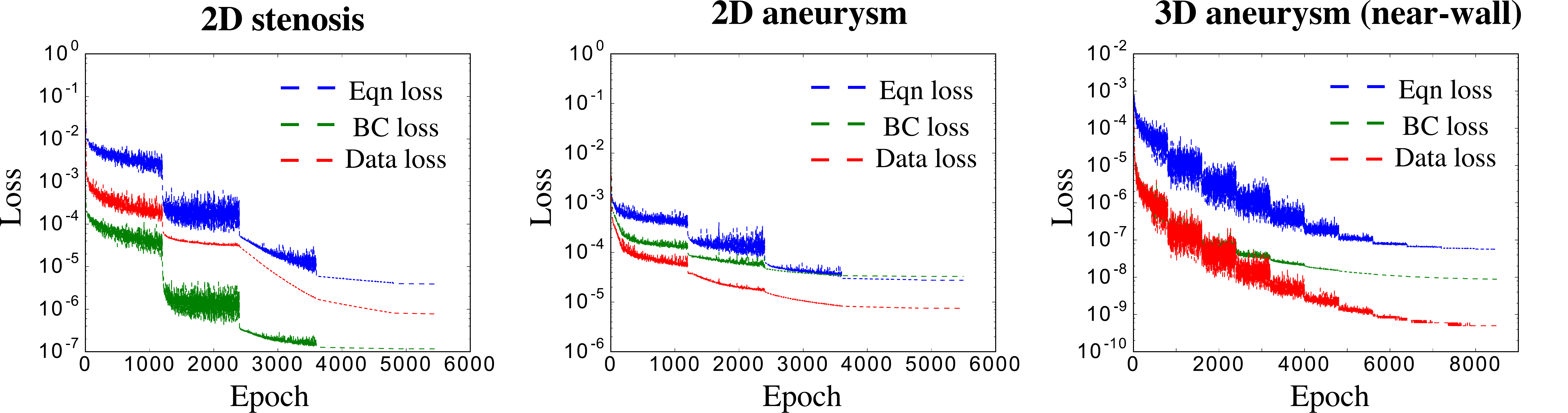}
\caption{The mean squared error (MSE) loss versus the epoch (iteration) number is plotted for the 2D stenosis, 2D aneurysm, and 3D aneurysm models. The equation loss, boundary condition loss, and data loss are plotted. The total loss is composed of a weighted sum of these losses. The loss in the 3D model is based on the near-wall region where PINN modeling was performed. } 
\label{fig:loss}
\end{figure}

The convergence of the 2D and 3D models is shown in Fig.~\ref{fig:loss} where the MSE losses are plotted. It is seen that the BC and data losses are consistently lower than the equation loss. The 3D model is based on the near-wall region, and therefore lower equation loss is observed in this model compared to the 2D models. 

\subsection{Test case 5: Parameter identification (viscosity)}

\begin{figure}[h!]
\centering
\includegraphics[scale =0.4]{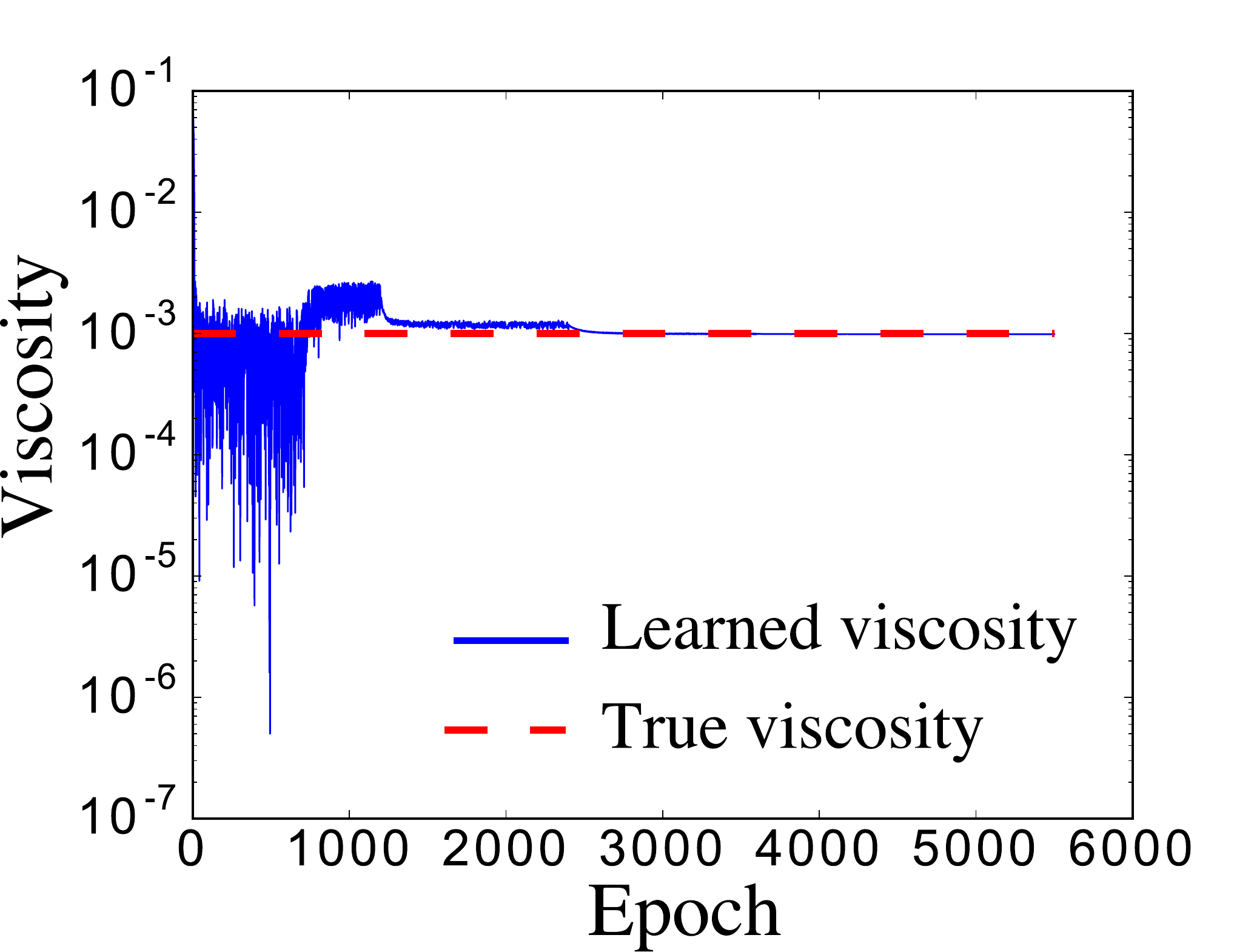}
\caption{Identification of viscosity using PINN is shown \edit{for the stenosis model.} The learned viscosity by PINN is plotted versus the number of epochs (iterations). PINN converges to the ground truth viscosity ($\mu=0.001$) after sufficient epochs. } 
\label{fig:mu}
\end{figure}

Finally, the viscosity identification results are shown in Fig.~\ref{fig:mu}. The plot shows the identified viscosity by PINN during different iterations. It could be seen that after a sufficient number of iterations, PINN successfully converges to the ground truth viscosity ($\mu=0.001$).

\begin{figure}[h!]
\centering
\includegraphics[scale =0.4]{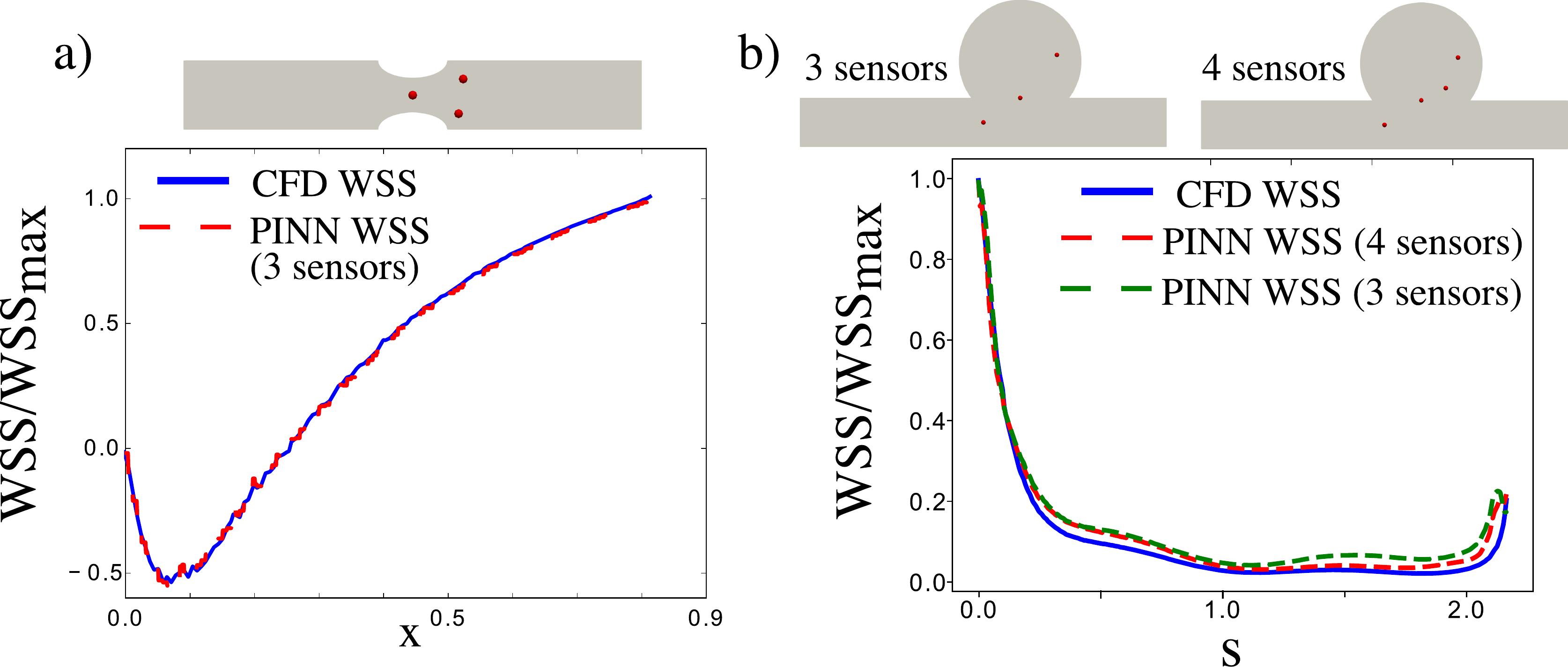}

\caption{\edit{The effect of reducing the number of sensors on the 2D results is shown. The sensor locations are shown with red dots. a) Three sensors are used in the 2D stenosis model and the predicted WSS results are compared with the reference CFD data. b) Four and three sensors are used in the 2D aneurysm model and the predicted WSS results are compared with the reference CFD data. } }
\label{fig:loc}
\end{figure}

\edit{
\subsection{The effect of sensor numbers}

The effect of reducing the number of sensors on WSS prediction in the 2D models is shown in Fig.~\ref{fig:loc}. In Fig.~\ref{fig:loc}a, three sensors are used in obtaining WSS data in the 2D stenosis model. The predicted value remains very close to the reference CFD data. The results with four sensors were very close to the three-sensor case and are therefore not shown. In Fig.~\ref{fig:loc}b, four and three sensors are used to predict WSS in the 2D aneurysm model where the reduction in accuracy could be seen when the number of sensors is reduced. 

}

\section{Discussion} \label{sec:disc}
In this manuscript, we proposed a PINN framework for estimating near-wall blood flow and WSS from sparse velocity measurements localized in a region of interest. The inlet and outlet boundary conditions were assumed unknown, thus representing challenges for traditional numerical methods. Such problems fall under the class of state estimation problems in fluid mechanics where the flow field is predicted with a limited number of point-wise measurements~\cite{NairGoza20}.

Our method's estimation of WSS from limited measurements was very accurate, where the model's prediction in 2D was indistinguishable from the ground truth data and in 3D the error was very small. The 1D transport problem proves the feasibility of our general framework. Namely, the analytical solution for the 1D problem (Eq.~\ref{eqn:analy}) shows that two unknown parameters ($c_1$ and $c_2$) need to be estimated when the equation and parameters are known but the BCs are not specified. Therefore, the sparse measurement data (three points in this example) define a regression problem that could be solved to obtain these coefficients. Of course, the sharp boundary layer in this example makes the regression problem ill-conditioned and it is necessary to select one point inside the boundary layer to overcome this issue. In the 2D problems, PINN did not correctly identify the inlet BC, however, the flow field in the region of interest where sensors were placed as well as the downstream region (in the stenosis problem) was correctly reconstructed. To explain this observation our PINN framework should be perceived as an inverse modeling problem where the inlet BC is to be determined. The issue with many inverse problems is that they usually do not have a unique solution. For example, different inlet velocity profiles with the same flow rate could give the same solution at a sufficiently downstream region. Therefore, to correctly identify the inlet BC one needs to place sensors in part of the domain where the flow is being developed. Our results showed that to get WSS in the region of interest we do not necessarily need to correctly identify the inlet velocity profile. 

The success in obtaining WSS from core flow measurement is tightly related to flow physics and the relation between WSS patterns and coherent structures away from the wall where the topological ~\cite{ArzaniShadden18} and dynamical~\cite{HabibiDawsonArzani20} features in WSS depend on flow patterns away from the wall. Velocity vectors away from the wall could be obtained theoretically using Taylor series expansion of wall quantities such as WSS and pressure~\cite{PerryChong86,GambarutoDoorlyYamaguchi10}. Machine learning models have been designed leveraging these mathematical relationships in estimating near-wall turbulent flows~\cite{MilanoKoumoutsakos02}. It is possible to account for these relationships in PINN when the data analysis is only focused on the near-wall region (e.g.,\ the 3D example in this study). 

While machine learning and deep learning are typically associated with large datasets, it is not clear if current 3D hemodynamics databases could provide generalizable information beyond very specific variations. The issue lies in the large landscape of all possible solutions once we realize the vast possibilities for variations in geometry and boundary conditions. What makes such purely data-driven approaches for estimating hemodynamics from morphology and boundary conditions~\cite{Suetal20,Feigeretal20} even more challenging is the high dimensionality of parameters of interest (e.g., WSS), which are typically vectorial quantities with spatiotemporal variation. Therefore, the notion of ``smart data'' compared to ``large data''~\cite{Bruntonetal20} is expected to be the new frontier in hemodynamics machine learning research. Sparsity based models have high potential in hemodynamics modeling (reviewed in~\cite{ArzaniDawson21}) and the present work demonstrates the power of physics-informed machine learning in leveraging sparse data for near-wall blood flow modeling.

The concept of a digital twin technology in modeling the cardiovascular system has recently gained attention~\cite{Chakshuetal21}. A digital twin can provide a real-time virtual representation for the cardiovascular system for the purpose of modeling. Digital twins often rely on a continuous feed of data. However, obtaining real-time large datasets and processing them in real-time represents a major challenge~\cite{Bruntonetal20}. Physics-informed machine learning models  could eliminate the need for large datasets and therefore are expected to be a key component in future digital twin models. While in our study we arbitrarily selected our sparse measurement data, in more practical problems, optimal sensor placement and data collection need to be addressed to improve model efficiency and even further reduce the amount of data needed~\cite{Manoharetal18,ArzaniDawson21,DengHeLiu21}.

Our study could be extended in multiple directions. Convolutional neural networks could be used in place of a fully connected neural network to improve PINN's convergence and accuracy by leveraging the strong spatial dependence in hemodynamics data. Traditional convolutional neural networks are designed for data defined on rectangular spaces (similar to a picture) and extension to complex geometries is an active research area~\cite{GaoSunWang21}. An optimal selection of hyperparameters in PINN is another area of investigation. For instance, in this work, the weight parameters $ \lambda_{b}$ and $ \lambda_{d}$ were selected based on trial and error. Recent work suggests an adaptive selection of these parameters based on neural tangent kernels~\cite{WangYuPerdikaris20}.  In real-world examples, outliers in input measurement data could exist~\cite{Wangetal20} that need to be detected. Noise in measurement data is another practical challenge.  It is possible that different measurement points have a different signal-to-noise ratio and therefore the input data could be of multi-fidelity nature. We could weight the input data in PINN's data loss based on estimates of data quality and noise. Rigorous approaches such as Bayesian PINN~\cite{SunWang20} and multi-fidelity PINN~\cite{MengKarniadakis20} have been proposed to handle such datasets and consider data uncertainty in a probabilistic way. \edit{Our presented model has some limitations.} Patient-specific geometries and pulsatile blood flow as well as a computationally efficient PINN framework to handle these complex blood flow environments should be investigated in future work.  \edit{We did not rigorously investigate optimal sensor placement, which is an interesting topic for future investigation. Finally, the success of the method with real-world experimental data remains to be investigated}

\section{Conclusion}

In this study, PINN was used to obtain  near-wall hemodynamics and WSS data from sparse velocity measurements and without knowledge of the inlet/outlet boundary conditions. We evaluated our framework in 1D advection-diffusion transport, 2D blood flow in a stenosis, 2D blood flow in an aneurysm, and 3D blood flow in an idealized aneurysm. WSS was obtained with very high accuracy. The proposed scientific machine learning framework demonstrates that partial knowledge about the physics of cardiovascular flows (e.g, Navier-Stokes equations without boundary conditions) could be combined with sparse data collected in a localized region to obtain hemodynamic parameters of clinical interest such as WSS.

\section*{Conflict of interest}
The authors declare no conflict of interest.

\section*{Acknowledgement}
The first author would like to thank fruitful discussions with Dr. Ricardo Vinuesa,  Dr. Scott Dawson, and first author's graduate students regarding the present work. JXW gratefully acknowledge the support from the National Science Foundation (CMMI-1934300).


\section*{Data Availability}
\edit{
The Pytorch codes and data used to generate the results in the manuscript are available on Github \url{https://github.com/amir-cardiolab/PINN-wss}.
}

\bibliographystyle{unsrt}
\bibliography{data_PINN}

\begin{thebibliography}{10}

\bibitem{VardhanRandles21}
M~Vardhan and A~Randles.
\newblock Application of physics-based flow models in cardiovascular medicine:
  Current practices and challenges.
\newblock {\em Biophysics Reviews}, 2(1):011302, 2021.

\bibitem{Arzanietal16}
A.~Arzani, A.~M. Gambaruto, G.~Chen, and S.~C. Shadden.
\newblock Lagrangian wall shear stress structures and near-wall transport in
  high-{Schmidt}-number aneurysmal flows.
\newblock {\em Journal of Fluid Mechanics}, 790:158--172, 2016.

\bibitem{Mahmoudietal20}
M.~Mahmoudi, A.~Farghadan, D.~R. McConnell, A.~J. Barker, J.~J. Wentzel, M.~J.
  Budoff, and A.~Arzani.
\newblock The story of wall shear stress in coronary artery atherosclerosis:
  biochemical transport and mechanotransduction.
\newblock {\em Journal of Biomechanical Engineering}, 143(4), 2020.

\bibitem{Samadyetal11}
H.~Samady, P.~Eshtehardi, M.~C. McDaniel, J.~Suo, S.~S. Dhawan, C.~Maynard,
  L.~H. Timmins, A.~A. Quyyumi, and D.~P. Giddens.
\newblock Coronary artery wall shear stress is associated with progression and
  transformation of atherosclerotic plaque and arterial remodeling in patients
  with coronary artery disease.
\newblock {\em Circulation}, 124(7):779--788, 2011.

\bibitem{Detmeretal20}
F.~J. Detmer, D.~L{\"u}ckehe, F.~Mut, M.~Slawski, S.~Hirsch, P.~Bijlenga,
  G.~von Voigt, and J.~R. Cebral.
\newblock Comparison of statistical learning approaches for cerebral aneurysm
  rupture assessment.
\newblock {\em International Journal of Computer Assisted Radiology and
  Surgery}, 15(1):141--150, 2020.

\bibitem{ArzaniShadden12}
A.~Arzani and S.~C. Shadden.
\newblock Characterization of the transport topology in patient-specific
  abdominal aortic aneurysm models.
\newblock {\em Physics of Fluids}, 24(8):1901, 2012.

\bibitem{AntigaSteinma09}
L.~Antiga and D.~A. Steinman.
\newblock Rethinking turbulence in blood.
\newblock {\em Biorheology}, 46(2):77--81, 2009.

\bibitem{ArzaniShadden18}
A.~Arzani and S.~C. Shadden.
\newblock Wall shear stress fixed points in cardiovascular fluid mechanics.
\newblock {\em Journal of Biomechanics}, 73:145--152, 2018.

\bibitem{ArzaniShadden16}
A.~Arzani and S.~C. Shadden.
\newblock Characterizations and correlations of wall shear stress in aneurysmal
  flow.
\newblock {\em Journal of Biomechanical Engineering}, 138(1):014503, 2016.

\bibitem{Mazzietal21}
V.~Mazzi, U.~Morbiducci, K.~Cal{\`o}, G.~De~Nisco, M.~Lodi~Rizzini, E.~Torta,
  G.~C.~A. Caridi, C.~Chiastra, and D.~Gallo.
\newblock Wall shear stress topological skeleton analysis in cardiovascular
  flows: Methods and applications.
\newblock {\em Mathematics}, 9(7):720, 2021.

\bibitem{Raffeletal18}
M.~Raffel, C.~E. Willert, F.~Scarano, C.~J. K{\"a}hler, S.~T. Wereley, and
  J.~Kompenhans.
\newblock {\em Particle image velocimetry: a practical guide}.
\newblock Springer, 2018.

\bibitem{Wangetal20}
H.~Wang, Z.~Yang, B.~Li, and S.~Wang.
\newblock Predicting the near-wall velocity of wall turbulence using a neural
  network for particle image velocimetry.
\newblock {\em Physics of Fluids}, 32(11):115105, 2020.

\bibitem{ArzaniDawson21}
A.~Arzani and S.~Dawson.
\newblock Data-driven cardiovascular flow modelling: examples and
  opportunities.
\newblock {\em Journal of The Royal Society Interface}, 18:20200802, 2021.

\bibitem{Bakhshinejadetal17}
A.~Bakhshinejad, A.~Baghaie, A.~Vali, D.~Saloner, V.~L. Rayz, and R.~M.
  D'Souza.
\newblock Merging computational fluid dynamics and {4D Flow MRI} using proper
  orthogonal decomposition and ridge regression.
\newblock {\em Journal of Biomechanics}, 58:162--173, 2017.

\bibitem{Gaidziketal19}
F.~Gaidzik, D.~Stucht, C.~Roloff, O.~Speck, D.~Th{\'e}venin, and G.~Janiga.
\newblock Transient flow prediction in an idealized aneurysm geometry using
  data assimilation.
\newblock {\em Computers in Biology and Medicine}, 115:103507, 2019.

\bibitem{Funkeetal19}
S.~W. Funke, M.~Nordaas, {\O}.~Evju, M.~S. Aln{\ae}s, and K.~A. Mardal.
\newblock Variational data assimilation for transient blood flow simulations:
  Cerebral aneurysms as an illustrative example.
\newblock {\em International Journal for Numerical Methods in Biomedical
  Engineering}, 35(1):e3152, 2019.

\bibitem{Habibietal21}
M.~Habibi, R.~M. D'Souza, S.~Dawson, and A.~Arzani.
\newblock Integrating multi-fidelity blood flow data with reduced-order data
  assimilation.
\newblock {\em arXiv preprint arXiv:2104.01971}, 2021.

\bibitem{GaoWang21}
H.~Gao and J.~X. Wang.
\newblock A bi-fidelity ensemble {Kalman} method for {PDE-constrained} inverse
  problems in computational mechanics.
\newblock {\em Computational Mechanics}, pages 1--17, 2021.

\bibitem{Raissietal19}
M.~Raissi, P.~Perdikaris, and G.~E. Karniadakis.
\newblock Physics-informed neural networks: A deep learning framework for
  solving forward and inverse problems involving nonlinear partial differential
  equations.
\newblock {\em Journal of Computational Physics}, 378:686--707, 2019.

\bibitem{Horniketal89}
K.~Hornik, M.~Stinchcombe, and H.~White.
\newblock Multilayer feedforward networks are universal approximators.
\newblock {\em Neural networks}, 2(5):359--366, 1989.

\bibitem{LeeYou19}
S.~Lee and D.~You.
\newblock Data-driven prediction of unsteady flow over a circular cylinder
  using deep learning.
\newblock {\em Journal of Fluid Mechanics}, 879:217--254, 2019.

\bibitem{Wesselsetal20}
H.~Wessels, C.~Wei{\ss}enfels, and P.~Wriggers.
\newblock The neural particle method--an updated {Lagrangian} physics informed
  neural network for computational fluid dynamics.
\newblock {\em Computer Methods in Applied Mechanics and Engineering},
  368:113127, 2020.

\bibitem{Kashinathetal21}
K.~Kashinath, M.~Mustafa, A.~Albert, J.~L. Wu, C.~Jiang, et~al.
\newblock Physics-informed machine learning: case studies for weather and
  climate modelling.
\newblock {\em Philosophical Transactions of the Royal Society A},
  379(2194):20200093, 2021.

\bibitem{Lucoretal21}
D.~Lucor, A.~Agrawal, and A.~Sergent.
\newblock Physics-aware deep neural networks for surrogate modeling of
  turbulent natural convection.
\newblock {\em arXiv preprint arXiv:2103.03565}, 2021.

\bibitem{GasmiTchelepi21}
C.~F. Gasmi and H.~Tchelepi.
\newblock Physics informed deep learning for flow and transport in porous
  media.
\newblock {\em arXiv preprint arXiv:2104.02629}, 2021.

\bibitem{Jinetal21}
X.~Jin, S.~Cai, H.~Li, and G.~E. Karniadakis.
\newblock {NSFnets (Navier-Stokes flow nets)}: Physics-informed neural networks
  for the incompressible {Navier-Stokes} equations.
\newblock {\em Journal of Computational Physics}, 426:109951, 2021.

\bibitem{Luetal21}
L.~Lu, X.~Meng, Z.~Mao, and G.~E. Karniadakis.
\newblock {DeepXDE}: A deep learning library for solving differential
  equations.
\newblock {\em SIAM Review}, 63(1):208--228, 2021.

\bibitem{Hennighetal20}
O.~Hennigh, S.~Narasimhan, M.~A. Nabian, A.~Subramaniam, K.~Tangsali,
  M.~Rietmann, et~al.
\newblock Nvidia simnet\^{}$\{$TM$\}$: an {AI-accelerated} multi-physics
  simulation framework.
\newblock {\em arXiv preprint arXiv:2012.07938}, 2020.

\bibitem{Sunetal20}
L.~Sun, H.~Gao, S.~Pan, and J.~X. Wang.
\newblock Surrogate modeling for fluid flows based on physics-constrained deep
  learning without simulation data.
\newblock {\em Computer Methods in Applied Mechanics and Engineering},
  361:112732, 2020.

\bibitem{Raissietal20}
M.~Raissi, A.~Yazdani, and G.~E. Karniadakis.
\newblock Hidden fluid mechanics: Learning velocity and pressure fields from
  flow visualizations.
\newblock {\em Science}, 367(6481):1026--1030, 2020.

\bibitem{Caietal21}
S.~Cai, H.~Li, F.~Zheng, F.~Kong, M.~Dao, G.~E. Karniadakis, and S.~Suresh.
\newblock Artificial intelligence velocimetry and microaneurysm-on-a-chip for
  three-dimensional analysis of blood flow in physiology and disease.
\newblock {\em Proceedings of the National Academy of Sciences}, 118(13), 2021.

\bibitem{Kissasetal20}
G.~Kissas, Y.~Yang, E.~Hwuang, W.~R. Witschey, J.~A. Detre, and P.~Perdikaris.
\newblock Machine learning in cardiovascular flows modeling: Predicting
  arterial blood pressure from non-invasive {4D flow MRI} data using
  physics-informed neural networks.
\newblock {\em Computer Methods in Applied Mechanics and Engineering},
  358:112623, 2020.

\bibitem{Reyesetal20}
B.~Reyes, A.~A. Howard, P.~Perdikaris, and A.~M. Tartakovsky.
\newblock Learning unknown physics of {non-Newtonian} fluids.
\newblock {\em arXiv preprint arXiv:2009.01658}, 2020.

\bibitem{LiuLiangSun20}
M.~Liu, L.~Liang, and W.~Sun.
\newblock A generic physics-informed neural network-based constitutive model
  for soft biological tissues.
\newblock {\em Computer Methods in Applied Mechanics and Engineering},
  372:113402, 2020.

\bibitem{Buosoetal21}
S.~Buoso, T.~Joyce, and S.~Kozerke.
\newblock Personalising left-ventricular biophysical models of the heart using
  parametric physics-informed neural networks.
\newblock {\em Medical Image Analysis}, page 102066, 2021.

\bibitem{Liuetal20}
B.~Liu, J.~Tang, H.~Huang, and X.~Y. Lu.
\newblock Deep learning methods for super-resolution reconstruction of
  turbulent flows.
\newblock {\em Physics of Fluids}, 32(2):025105, 2020.

\bibitem{Guemesetal21}
A.~G{\"u}emes, H.~Tober, S.~Discetti, A.~Ianiro, B.~Sirmacek, H.~Azizpour, and
  R.~Vinuesa.
\newblock From coarse wall measurements to turbulent velocity fields with deep
  learning.
\newblock {\em arXiv preprint arXiv:2103.07387}, 2021.

\bibitem{Fukamietal21}
K.~Fukami, K.~Fukagata, and K.~Taira.
\newblock Machine-learning-based spatio-temporal super resolution
  reconstruction of turbulent flows.
\newblock {\em Journal of Fluid Mechanics}, 909, 2021.

\bibitem{GaoSunWang20}
H.~Gao, L.~Sun, and J.~X. Wang.
\newblock Super-resolution and denoising of fluid flow using physics-informed
  convolutional neural networks without high-resolution labels.
\newblock {\em arXiv preprint arXiv:2011.02364}, 2020.

\bibitem{Fathietal20}
M.~F. Fathi, I.~Perez-Raya, A.~Baghaie, P.~Berg, G.~Janiga, A.~Arzani, and
  R.~M. D'Souza.
\newblock Super-resolution and denoising of {4D-Flow MRI} using
  physics-informed deep neural nets.
\newblock {\em Computer Methods and Programs in Biomedicine}, page 105729,
  2020.

\bibitem{Guastonietal20}
L.~Guastoni, M.~P. Encinar, P.~Schlatter, H.~Azizpour, and R.~Vinuesa.
\newblock Prediction of wall-bounded turbulence from wall quantities using
  convolutional neural networks.
\newblock In {\em Journal of Physics: Conference Series}, volume 1522, page
  012022, 2020.

\bibitem{Pengetal20}
G.~C.~Y. Peng, M.~Alber, A.~B. Tepole, W.~R. Cannon, S.~De, S.~Dura-Bernal,
  K.~Garikipati, G.~Karniadakis, W.~W. Lytton, P.~Perdikaris, L.~Petzold, and
  E.~Kuhl.
\newblock Multiscale modeling meets machine learning: What can we learn?
\newblock {\em Archives of Computational Methods in Engineering}, pages 1--21,
  2020.

\bibitem{Ramachandranetal17}
P.~Ramachandran, B.~Zoph, and Q.~V. Le.
\newblock Searching for activation functions.
\newblock {\em arXiv preprint arXiv:1710.05941}, 2017.

\bibitem{LoggMardalWells12}
A.~Logg, K.~A. Mardal, and G.~Wells.
\newblock {\em Automated solution of differential equations by the finite
  element method}, volume~84.
\newblock Springer, Berlin, Heidelberg, 2012.

\bibitem{LeidermanFogelson11}
K.~Leiderman and A.~L. Fogelson.
\newblock Grow with the flow: a spatial--temporal model of platelet deposition
  and blood coagulation under flow.
\newblock {\em Mathematical Medicine and Biology}, 28(1):47--84, 2011.

\bibitem{HansenArzaniShadden19}
K.~B. Hansen, A.~Arzani, and S.~C. Shadden.
\newblock Finite element modeling of near-wall mass transport in cardiovascular
  flows.
\newblock {\em International Journal for Numerical Methods in Biomedical
  Engineering}, 35(1):e3148, 2019.

\bibitem{ShaddenHendabadi13}
S.~C. Shadden and S.~Hendabadi.
\newblock Potential fluid mechanic pathways of platelet activation.
\newblock {\em Biomechanics and Modeling in Mechanobiology}, 12(3):467--474,
  2013.

\bibitem{Arzani20}
A.~Arzani.
\newblock Coronary artery plaque growth: A two-way coupled shear stress--driven
  model.
\newblock {\em International Journal for Numerical Methods in Biomedical
  Engineering}, 36(1):e3293, 2020.

\bibitem{Baeketal10}
H.~Baek, M.~V. Jayaraman, P.~D. Richardson, and G.~E. Karniadakis.
\newblock Flow instability and wall shear stress variation in intracranial
  aneurysms.
\newblock {\em Journal of the Royal Society Interface}, 7(47):967--988, 2010.

\bibitem{Cebraletal11}
J.~R. Cebral, F.~Mut, J.~Weir, and C.~M. Putman.
\newblock Association of hemodynamic characteristics and cerebral aneurysm
  rupture.
\newblock {\em American Journal of Neuroradiology}, 32(2):264--270, 2011.

\bibitem{ChoMooneyCho08}
Y.~I. Cho, M.~P. Mooney, and D.~J. Cho.
\newblock Hemorheological disorders in diabetes mellitus.
\newblock {\em Journal of Diabetes Science and Technology}, 2(6):1130--1138,
  2008.

\bibitem{Arzani18}
A.~Arzani.
\newblock Accounting for residence-time in blood rheology models: do we really
  need {non-Newtonian} blood flow modelling in large arteries?
\newblock {\em Journal of The Royal Society Interface}, 15(146):20180486, 2018.

\bibitem{NairGoza20}
N.~J. Nair and A.~Goza.
\newblock Leveraging reduced-order models for state estimation using deep
  learning.
\newblock {\em Journal of Fluid Mechanics}, 897, 2020.

\bibitem{HabibiDawsonArzani20}
M.~Habibi, S.T.M. Dawson, and A.~Arzani.
\newblock Data-driven pulsatile blood flow physics with dynamic mode
  decomposition.
\newblock {\em Fluids}, 5(3):111, 2020.

\bibitem{PerryChong86}
A.~E. Perry and M.~S. Chong.
\newblock A series-expansion study of the {Navier--Stokes} equations with
  applications to three-dimensional separation patterns.
\newblock {\em Journal of Fluid Mechanics}, 173:207--223, 1986.

\bibitem{GambarutoDoorlyYamaguchi10}
A.~M. Gambaruto, D.~J. Doorly, and T.~Yamaguchi.
\newblock Wall shear stress and near-wall convective transport: Comparisons
  with vascular remodelling in a peripheral graft anastomosis.
\newblock {\em Journal of Computational Physics}, 229(14):5339--5356, 2010.

\bibitem{MilanoKoumoutsakos02}
M.~Milano and P.~Koumoutsakos.
\newblock Neural network modeling for near wall turbulent flow.
\newblock {\em Journal of Computational Physics}, 182(1):1--26, 2002.

\bibitem{Suetal20}
B.~Su, J.~M. Zhang, H.~Zou, D.~Ghista, T.~T. Le, and C.~Chin.
\newblock Generating wall shear stress for coronary artery in real-time using
  neural networks: Feasibility and initial results based on idealized models.
\newblock {\em Computers in Biology and Medicine}, 126:104038, 2020.

\bibitem{Feigeretal20}
B.~Feiger, J.~Gounley, D.~Adler, J.~A. Leopold, E.~W. Draeger, R.~Chaudhury,
  J.~Ryan, G.~Pathangey, K.~Winarta, D.~Frakes, F.~Michor, and A.~Randles.
\newblock Accelerating massively parallel hemodynamic models of coarctation of
  the aorta using neural networks.
\newblock {\em Scientific Reports}, 10(1):1--13, 2020.

\bibitem{Bruntonetal20}
S.~L. Brunton, J.~N. Kutz, K.~Manohar, A.~Y. Aravkin, et~al.
\newblock Data-driven aerospace engineering: Reframing the industry with
  machine learning.
\newblock {\em arXiv preprint arXiv:2008.10740}, 2020.

\bibitem{Chakshuetal21}
N.~K. Chakshu, I.~Sazonov, and P.~Nithiarasu.
\newblock Towards enabling a cardiovascular digital twin for human systemic
  circulation using inverse analysis.
\newblock {\em Biomechanics and Modeling in Mechanobiology}, 20(2):449--465,
  2021.

\bibitem{Manoharetal18}
K.~Manohar, B.~W. Brunton, J.~N. Kutz, and S.~L. Brunton.
\newblock Data-driven sparse sensor placement for reconstruction: Demonstrating
  the benefits of exploiting known patterns.
\newblock {\em IEEE Control Systems Magazine}, 38(3):63--86, 2018.

\bibitem{DengHeLiu21}
Z.~Deng, C.~He, and Y.~Liu.
\newblock Deep neural network-based strategy for optimal sensor placement in
  data assimilation of turbulent flow.
\newblock {\em Physics of Fluids}, 33(2):025119, 2021.

\bibitem{GaoSunWang21}
H.~Gao, L.~Sun, and J.~X. Wang.
\newblock {PhyGeoNet}: Physics-informed geometry-adaptive convolutional neural
  networks for solving parameterized steady-state {PDEs} on irregular domain.
\newblock {\em Journal of Computational Physics}, 428:110079, 2021.

\bibitem{WangYuPerdikaris20}
S.~Wang, X.~Yu, and P.~Perdikaris.
\newblock When and why {PINNs} fail to train: A neural tangent kernel
  perspective.
\newblock {\em arXiv preprint arXiv:2007.14527}, 2020.

\bibitem{SunWang20}
L.~Sun and J.~X. Wang.
\newblock Physics-constrained bayesian neural network for fluid flow
  reconstruction with sparse and noisy data.
\newblock {\em Theoretical and Applied Mechanics Letters}, 10(3):161--169,
  2020.

\bibitem{MengKarniadakis20}
X.~Meng and G.~E. Karniadakis.
\newblock A composite neural network that learns from multi-fidelity data:
  Application to function approximation and inverse {PDE} problems.
\newblock {\em Journal of Computational Physics}, 401:109020, 2020.

\end{thebibliography}

\end{document}